\documentclass[runningheads]{llncs}
\usepackage{graphicx}
\usepackage{graphics}
\usepackage{wrapfig}
\usepackage{float}
\usepackage{textcomp}
\usepackage{url}
\begin{document}
\title{Deep Quantitative Susceptibility Mapping for Background Field Removal and Total Field Inversion}
%
%

\author{{Juan Liu\inst{1,2}} \and
Kevin M. Koch\inst{1,2,3}}
\authorrunning{J. Liu, K.M. Koch}
%
\institute{Center for Imaging Research, Medical College of Wisconsin, Milwaukee, WI, USA \and
Biomedical Engineering, Marquette university and Medical College of Wisconsin, Milwaukee, WI, USA 
\and
Radiology, Medical College of Wisconsin, Milwaukee, WI, USA \\
\email{kmkoch@mcw.edu} }

\maketitle  
\begin{abstract}

Quantitative susceptibility mapping (QSM) utilizes MRI signal phase to estimate local tissue susceptibility, which has been shown useful to provide novel image contrast and as biomarkers of abnormal tissue. QSM requires addressing a challenging post-processing problem: filtering of image phase estimates and inversion of the phase to susceptibility relationship. A wide variety of quantification errors, robustness limitations, and artifacts constraints QSM clinical translation. To overcome these limitations, a robust deep-learning-based QSM reconstruction approach is proposed to perform background field removal and susceptibility inversion simultaneously from input MRI phase images. Synthetic training data based on in-vivo data sources and physics simulations were used for training. The network was quantitatively tested using gold-standard in-silico labeled dataset against established background field removal and QSM inversion approaches. In addition, the algorithm was applied to a QSM challenge data and clinical susceptibility-weighted imaging (SWI) data. When quantitatively compared against gold-standard in-silico labels, the proposed algorithm outperformed the existing comparable background field removal approaches and QSM reconstruction algorithms. The QSM challenge data and clinical SWI data demonstrated that the proposed approach was able to robustly generate high quality local field and QSM with improved accuracy. 

\end{abstract}

\newpage     


\section{Introduction}

Magnetic susceptibility has long been utilized to generate diagnostic imaging contrast through magnetic resonance imaging (MRI) techniques such as T$_2^*$ or enhanced susceptibility-weighted imaging (SWI)~\cite{haacke2004susceptibility}. Quantitative Susceptibility Mapping (QSM) is a MRI post-processing technique that expands upon these widely utilized susceptibility-weighted practices to provide estimates of underlying tissue magnetism~\cite{wang2015quantitative}. Existing studies have explored to usage of QSM to quantification of specific biomarkers\cite{haacke2005imaging,deistung2013toward,zhang2015quantitative,zheng2013measuring}, brain tumors\cite{deistung2013quantitative,wisnieff2015quantitative}, neurodegenerative disorders\cite{bilgic2012mri,wisnieff2015quantitative}, iron overload in the liver\cite{sharma2015quantitative}, blood oxygenation assessment\cite{fan2014quantitative}, and mild traumatic brain injury\cite{koch2018quantitative} as well.

QSM is performed by collecting phase-sensitive MRI to estimate local magnetic perturbations along the direction of the polarizing \textbf{B}$_0$ magnetic field~\cite{wang2015quantitative}. QSM transformation from raw perturbation fields usually consist of two steps: (i) background field removal to determine the local tissue field, and (ii) inversion from the local field to the tissue susceptibility. Both steps require solving ill-posed inverse computational problems, especially the field-to-source inversion, which pose challenges for accurate susceptibility quantification.

Background field removal (BFR) methods typically rely on spatial filtering and/or dipole field modelling methods~\cite{schweser2011quantitative,sun2014background,liu2011novel,zhou2014background}. Existing BFR approaches can often perform well under ideal circumstances with perfect brain tissue masking. However, they suffer from several constraints, such as brain mask erosion, parameter tuning, inaccurate background field removal close to brain boundary, and shading artifacts, which can result in local field estimation errors and subsequently susceptibility quantification errors. In clinical applications, large slice thickness and resulting non-isotropic acquisition resolutions further hinder the performance of existing background removal methods.

After background field removal step to retrieve the local field field, QSM inversion is performed to solve the field-to-source problem and estimate underlying tissue susceptibility distributions. Though clinically challenging, acquiring data at multiple orientations to the magnetic field remains the empirical gold-standard for \emph{in vivo} QSM assessment, as the additional field data sufficiently improves the conditioning of the inversion algorithm~\cite{liu2009calculation}. In the absence of multiple-orientation acquisitions, single-orientation susceptibility maps are computed by either thresholding of the convolution operator \cite{shmueli2009magnetic,wharton2010susceptibility,haacke2010susceptibility} or use of more sophisticated regularization methods \cite{de2008quantitative,liu2011morphology,bilgic2014fast}.

To avoid BFR error propagation into QSM estimation, several single-step QSM algorithms, which directly estimate the susceptibility distribution from the raw perturbation field, have been proposed. These approaches, hereby denoted as ``total field inversion" (TFI) methods, utilize algorithmic adjustments such as Laplacian spatial conditioning operators~\cite{Berkin2014SingleStep,Kristian2014SingleStep,Tian2014Differential,Samir2014SingleStep}, preconditioning boosted iterative inversions~\cite{liu2017preconditioned}, or total variation (TV) and/or Tikhonov regularized inversion~\cite{chatnuntawech2017single,sun2018whole} to address the challenges imposed by TFI. TFI methods have shown improved susceptibility estimation performance compared to conventional two-step QSM methods. However, all existing TFI methods require tuning of regularization parameters and long computation times. 

Both existing single-step and two-step QSM reconstruction approaches have limits. First, most of existing methods require tuning of regularization parameters, which can introduce introduce estimation biases and artifacts. Second, they require substantial computation times to solve their respected iterative optimization problems. Last but most important, these methods exhibit residual streaking artifacts in regions with large susceptibility variations, such as intracranial hemorrhages. Furthermore, existing available methods have compromised performance when applied to routinely acquired clinical susceptibility-weighted imaging data, which is highly non-isotropic in resolution (due to examination time constraints). 

To address the limitations of existing background field removal and QSM inversion methods, here we demonstrate a deep-learning-based approach to estimate local field and susceptibility maps at the same time. Leveraging the well-defined forward susceptibility-to-field physics model~\cite{marques2005application,salomir2003fast}, synthetic data are used for neural network training to overcome the bottleneck of collecting and curating accurate ground-truth training data. The proposed deep learning network is demonstrated on synthetic data, QSM challenge data, and clinical SWI data. Quantitative performance evaluations and several qualitative demonstrations of the proposed method against existing BFR approaches and QSM reconstruction algorithms are presented to show the performance of the proposed method relative to existing approaches.

\section{Methods}

\subsection{Training Data} 

200 \emph{in vivo} QSM datasets were used as inputs to the simulated neural network training data generator. The resolution of this source data was 0.5x0.5x2.0 mm$^3$. From the four echo time images in each data set, QSM estimates were generated using the following existing tools: brain masking using SPM~\cite{ashburner2005unified}, BFR using the Regularization-enabled Sophisticated Harmonic Artifact Reduction on Phase data (RESHARP) \cite{sun2014background} method, and susceptibility inversion was performed using a previously developed Approximated Susceptibility through Parcellated Encoder-decoder Networks (ASPEN)~\cite{liu2019quantitative}.  

\begin{figure}[ht]
\begin{center}
\includegraphics[width=14cm]{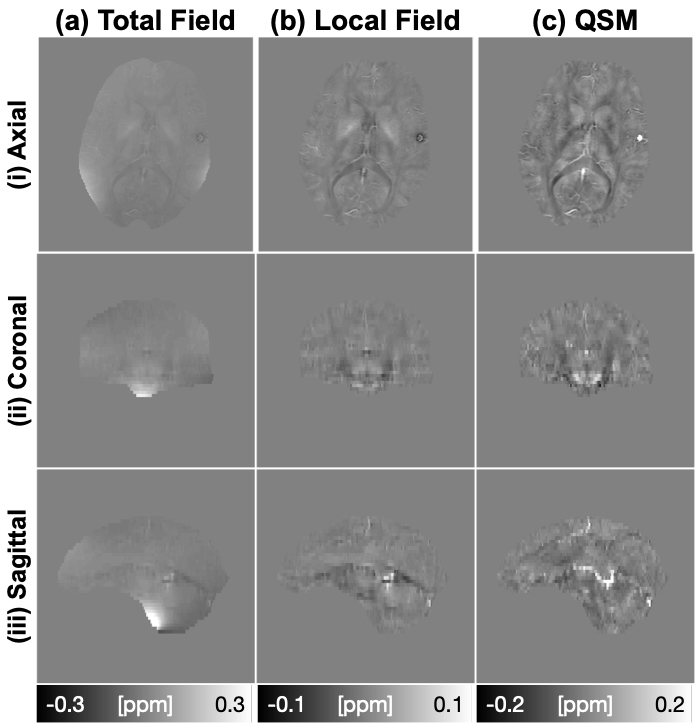}
\caption{Illustration of training dataset.
\label{fig_trainingdata} 
}
\end{center}
\end{figure}

Besides, geometric shapes such as ellipsoid, sphere, cuboid and cylinder with random susceptibility values and random orientations are randomly placed on the susceptibility maps to mimic hemorrhage and calcifications. Using the \emph{in-vivo} QSM estimates, local tissue magnetic perturbations were calculated using well-known dipole convolution methods~\cite{salomir2003fast,marques2005application}. The background perturbation fields were then simulated from random magnetic susceptibility sources to mimic background field. The field perturbation for training total field inversion was then constructed from the superimposed local tissue perturbations and the background perturbation. Therefore, the network is being trained exclusively by the ensuing closed-form forward source-to-field computation. Fig.~\ref{fig_trainingdata} provides an example from this training data construction process.

\subsection{Neural Network Architecture and Training} 

A 3D convolutional neural network with encoder-decoder architecture, based on a modified version of an established U-Net architecture \cite{ronneberger2015u}, was trained to perform whole brain pixel-wise susceptibility estimation and local tissue field, using whole brain total field and brain mask as the inputs. The encoder takes the inputs, and generates multiple levels features. The decoder decodes features at multiple levels aggregated by the encoder to generate a susceptibility distribution. Skip connections between the encoding path and decoding path were used to effectively transfer local feature information from the encoding path to the decoding path and facilitate faster training \cite{long2015fully,ronneberger2015u}. Gated convolution with LeakyReLU as activation function and Sigmoid for gating value was applied for automatically selecting adaptive features for each channel and each spatial location\cite{yu2018free}. Dilated gated convolution was applied in deeper convolutional layers to increase the receptive field. The last layer is a convolutional layer to output the local field and susceptibility maps.

For the image resolution 1.06 mm$^3$ isotropic, the neural network image size is set 160x160x160. For clinical data with image resolution 0.76x0.76x3.0 mm$^3$, the neural network input shape is 256x256x64. L1 loss and gradient difference loss between the local field and label was utilized as a loss function for local field loss. L1 loss and gradient difference loss between the susceptibility map and label was utilized as a loss function for susceptibility map loss. The total loss was the weighted sum of the two losses with equal loss weight. 6000 data were used for training. The RMSprop optimizer was used in the deep learning training. The initial learning rate was set as 0.0001, with exponential decay at every 200 steps. Three NVIDIA tesla k40 graphics processing units (GPUs) were used for training with batch size 3. The neural network was trained and evaluated using Keras with Tensorflow as backend.

\subsection{Performance Evaluation}

\subsubsection{Synthetic Data} 

100 simulated data sets with matrix size 160x160x160 and voxel size 1.06x1.06x1.06mm$^3$ generated in similar fashion to the training data without containing randomly inserted geometric shapes were used to evaluate the performance of QSMAllNet.

For background field removal evaluation, QSMAllNet was compared with existing background field removal methods, including (i) Sophisticated Harmonic Artifact Reduction on Phase data (SHARP)\cite{schweser2010differentiation}, (ii) Regularization-enabled SHARP \cite{sun2014background}, (iii) Projection onto Dipole Fields (PDF)\cite{liu2011novel}, (iv) Laplacian boundary value (LBV)\cite{zhou2014background}. 

For QSM estimation comparison, both single-step and two-steps QSM reconstruction approaches, were selected for comparison, including (i) single step total variation QSM (SS-TV-QSM) \cite{chatnuntawech2017single}, (ii) least-norm QSM (LN-QSM) \cite{sun2018whole}, (iii) RESHARP for BFR and Thresholded K-space Division (TKD) for QSM inversion for QSM inversion.

Estimation errors from each technique were computed using root mean squared error (RMSE), high-frequency error norm (HFEN), and structural similarity (SSIM) index with the ground truth local field and susceptibility maps. 

\subsubsection{2016 QSM Reconstruction Challenge Dataset} 

QSM reconstruction challenge dataset was acquired from a healthy 30-year-old female subject. Data for Calculation of Susceptibility through Multiple Orientation Sampling (COSMOS) and susceptibility-tensor computation was computed using a heavily accelerated wave-CAIPI~\cite{bilgic2015wave} acquisition at 1.06 mm isotropic resolution collected at 12 different head orientations at a single echo time~\cite{langkammer2018quantitative}. 

The provided background field was implemented using the LBV method after transmit phase removal by fitting and subtracting a fourth‐order 3D‐polynomial. To compare the background field removal performance of QSMAllNet, SHARP, RESHARP, PDF, LBV, and the proposed method were used to remove the background field using the unwrapped total field maps and brain mask provided.

For susceptibility estimates comparison, SS-TV-QSM, LN-QSM, TKD, Morphology Enabled Dipole Inversion (MEDI) \cite{liu2012morphology}, and QSMAllNet approaches were used to reconstruct QSM images. TKD and MEDI were performed using the provided local tissue field. TKD results were provided publically by the QSM challenge organizers, with threshold 0.19 which yields the best trade-off between quantification accuracy and artifacts. All methods were evaluated against the ``gold standard" Susceptibility Tensor Imaging (STI)~\cite{liu2010susceptibility} (3,3) component computational result provided with the challenge dataset~\cite{liu2010susceptibility}. 

\subsubsection{Clinical Data} 

One hundred clinical QSM data were acquired using gradient echo T$_2^*$ weighted angiography (SWAN, GE), a new method for SWI with short acquisition times, at a 3T MRI scanner (GE Healthcare MR750) with data acquisition parameters: in-plane data acquisition matrix 288x224, field of view 22 cm, slice thickness 3 mm, autocalibrated parallel imaging factors 1x2 or 1x3, number of slices 46-54, first echo time 12.6 ms, echo spacing 4.1 ms, number of echoes 7, flip angle 15$^o$, repetition time 39.7 ms, total scan time about 2 minutes.  

The SWI images were processed by vendor reconstruction algorithms. The raw k-space data were saved for offline QSM processing. Multi-echo real and imaginary data were reconstructed from k-space data, with reconstruction matrix size 288x288, voxel size 0.76x0.76x3.0 mm$^3$. The field map was obtained by the fitting of multi-echo phases. Brain masks were obtained using the FSL brain extraction tool.

Using the total field map and brain mask, representative state-of-the-arts methods of background field removal and QSM reconstruction, including SHARP, RESHARP, PDF, and LBV, SS-TV-QSM, LN-QSM, PDF+MEDI, and RESHARP+TKD were selected for comparison. Five clinical cases were chosen to demonstrate QSMAllNet performance. 

For the purposes of performance evaluation, for SHARP, a commonly used threshold of 0.05 was implemented. For SHARP and RESHARP, spherical kernel radius was set as 6mm to trade off the background removal performance and brain erosion; for TKD, the threshold was set to 0.20; for MEDI, the regularization factor was set to 1000; the TV and Tikhonov regularization parameters for LN-QSM were set to 4x10$^{-4}$ and 10$^{-3}$ respectively. MEDI toolbox, SS-TV-QSM code, LN-QSM code publicly provided by the authors were used to calculate the QSM images\cite{MEDI_cornell_WEB,berkin_MIT_WEB,LN_QSM_Code}.

\section{Results}

\subsubsection{Synthetic Data} 

Table.~\ref{table_sim_data_rdf} illustrates the RMSE, HFEN, and SSIM using five background field removal methods compared with ground truth from 100 synthetic test data. The proposed method achieved the best score in RMSE, HFEN, and SSIM. 

In Table.~\ref{table_sim_data_qsm}, the quantitative metrics of four QSM reconstruction methods compared with ground truth from 100 synthetic test data were shown. The proposed method achieved the best score in RMSE, HFEN, and SSIM. 

\begin{figure}[ht]
\begin{center}
\includegraphics[width=14cm]{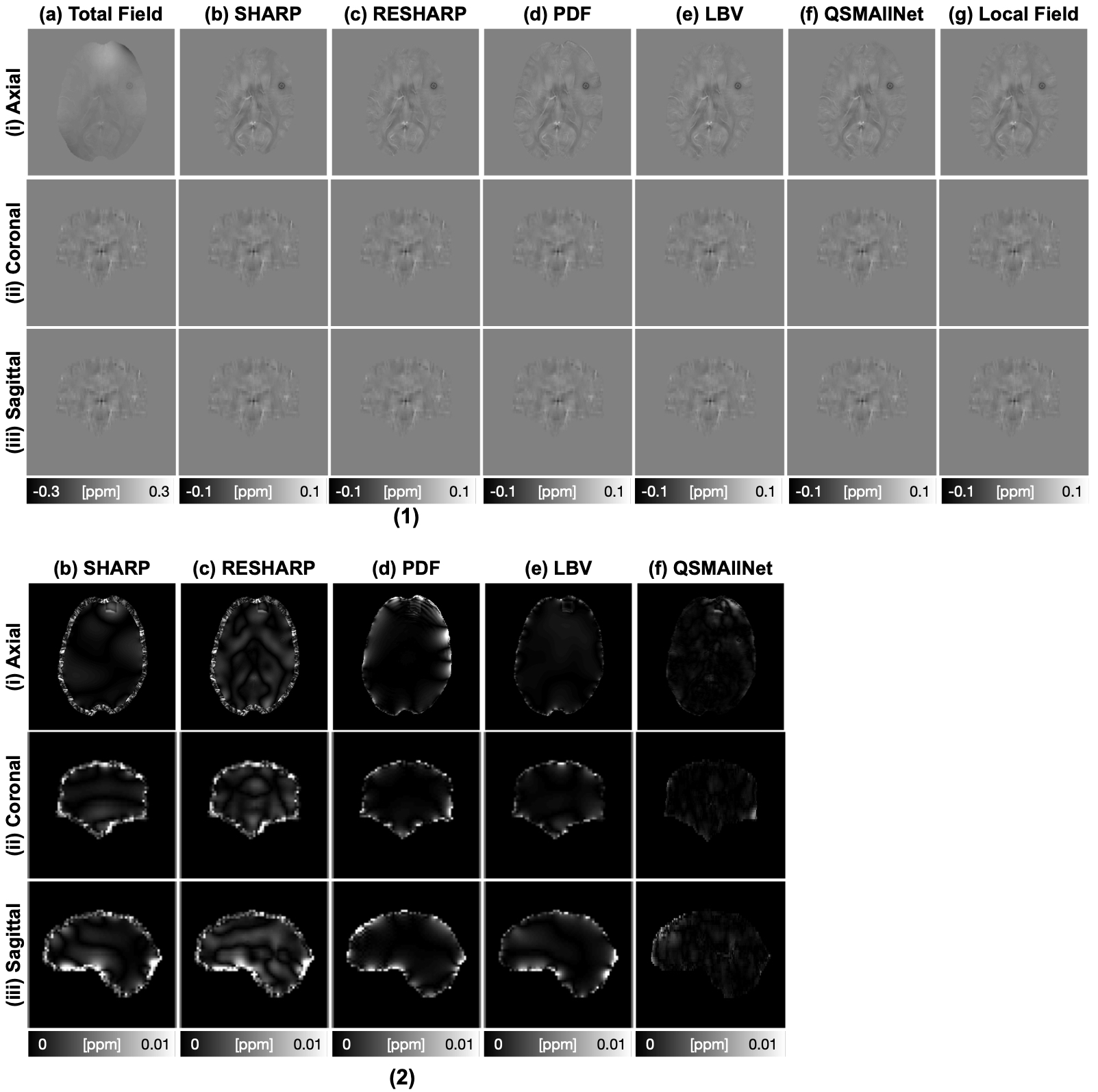}
\caption{Local field and residual error maps of one synthetic testing data. It is obviously shown that QSMAllNet have the least residual errors.  
\label{testdata_rdf}}
\end{center}
\end{figure}

Fig.\ref{testdata_rdf} (1) shows local field estimated by five different methods are compared with ground truth on one test data. The residual maps compared with ground truth are shown in Fig.\ref{testdata_rdf} (2). Based on visual assessment, QSMAllNet results have best local field estimation and least residual errors.  

\begin{figure}[ht]
\begin{center}
\includegraphics[width=14cm]{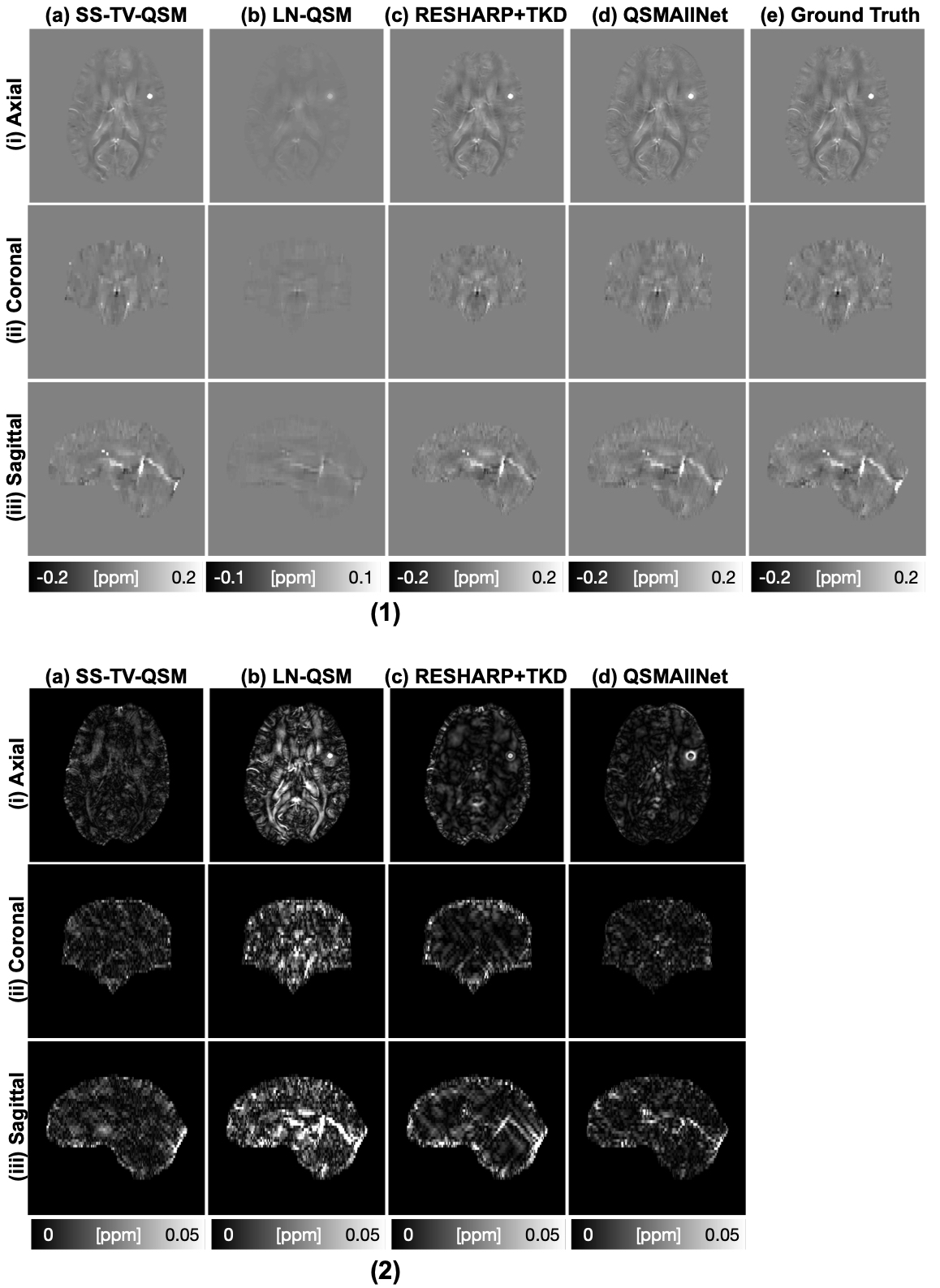}
\caption{QSM maps and residual error maps of one synthetic testing data.It is obviously shown that QSMAllNet have the least residual errors.  
\label{testdata_qsm}}
\end{center}
\end{figure}

In Fig.\ref{testdata_qsm} (1), QSM images reconstructed by four different reconstruction methods are compared with ground truth susceptibility maps on the test data in Fig.\ref{testdata_rdf}. The residual maps compared with ground truth are shown in Fig.\ref{testdata_qsm} (2). QSMAllNet results have best susceptibility estimations and least residual errors.  

\subsubsection{2016 QSM Reconstruction Challenge Dataset} 

\begin{figure}[ht]
\begin{center}
\includegraphics[width=14cm]{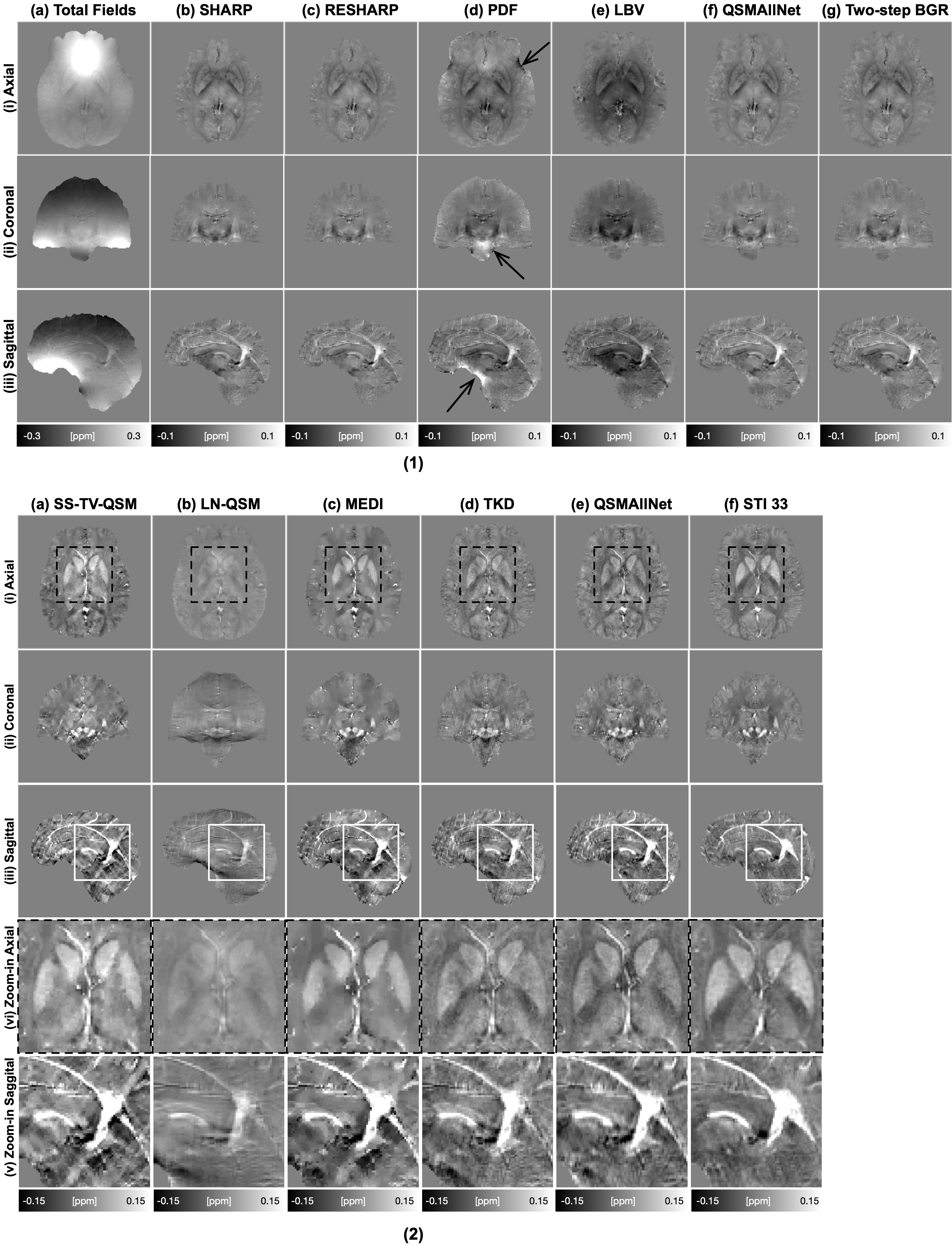}
\caption{Total fields and QSM background removal results (1), susceptibility images (2) on QSM challenge dataset. SHARP and RESHARP results (b-c) have brain erosion. LBV and PDF have shading artifacts in the tissue fields. PDF suffers from large inaccurate background field removal close to brain boundary (black arrows, 1, d, i-iii). QSMAllNet results show better background field removal. From QSM images, SS-TV-QS and MEDI suffers from image blurring and streaking artifacts. LN-QSM has intensity variation problems. TKD results show good image sharpness, but lose fine details. QSMAllNet shows improved susceptibility estimation with high image sharpness and nonvisible streaking artifacts.
\label{fig_qsmchallenge}}
\end{center}
\end{figure}

Fig.~\ref{fig_qsmchallenge} (1) illustrates total field map and local tissue fields from the 2016 ISMRM QSM Challenge dataset reconstructed using six methods, displayed in 3 reformatted planes(rows i, ii, iii). SHARP and RESHARP suffers from brain erosion (1, b-c, i-iii). PDF and LBV results show substantial shading artifacts (1, d-e, i-iii). It is clearly visible that PDF results have inaccurate background field removal close to brain boundary (black arrows, 1, d, i-iii). QSMAllNet shows high accuracy background field removal results and similar to the provided two-step local tissue field (1,g).

Fig.~\ref{fig_qsmchallenge} (2) illustrates susceptibility maps using five methods compared to the ground truth STI (3,3). Streaking artifacts are clearly identified in the sagittal reformat for SS-TV-QSM maps. In addition, the SS-TV-QSM map shows clearly compromised spatial resolution relative to the other maps. LN-QSM results suffer from image blurring and intensity variation across the image. MEDI results show substantial image blurring and streaking artifacts. TKD results show good image sharpness, but lose detail microstructures. These observations are amplified in the zoomed maps in the last two rows (iv, v), where clear performance improvements of the proposed method in reproducing the fine structure of the STI (3,3) map and no-visible streaking artifacts.

\subsubsection{Clinical Data}

\begin{figure}[ht]
\begin{center}
\includegraphics[width=14cm]{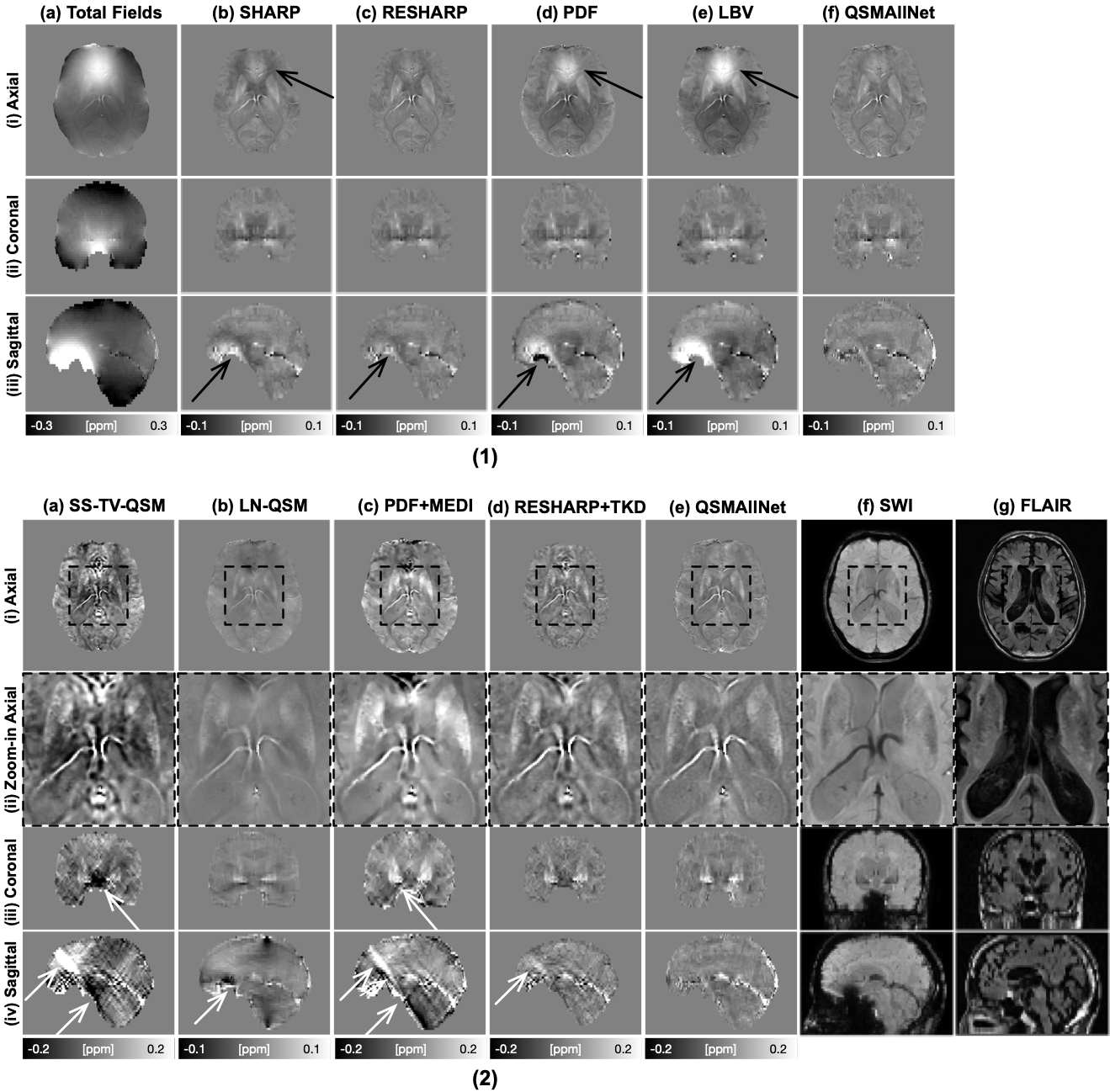}
\caption{Total fields and QSM background removal results (1), susceptibility images, SWI images and FLAIR images (2) on a 90-year-old patient with lung cancer. In (1), residual background fields are clearly visible in SHARP, RESHARP, PDF and LBV (b-d) results in axial and saggital views (black arrows). SHARP and RESHARP results (b-c) have brain erosion. LBV and PDF have shading artifacts in the tissue fields. QSMAllNet results show better background field removal. From the susceptibility maps (2), SS-TV-QSM, LN-QSM, PDF+MEDI, and RESHARP+TKD have susceptibility large estimation errors, especially streaking artifacts and shading artifacts (white arrows). Based on visual comparison, QSMAllNet can produce improved local field and susceptibility estimation. 
\label{p626}}
\end{center}
\end{figure}

In Fig.~\ref{p626}, background removal results and susceptibility images on a 90-year-old patient with lung cancer are shown in three (axial/coronal/sagittal) views. In (1), residual background fields are clearly visible in SHARP, RESHARP, PDF and LBV (1, b-d) results in axial and saggital views. SHARP and RESHARP results (1, b-c) have brain erosion. LBV and PDF have shading artifacts in the tissue fields. QSMAllNet results show better background field removal. From the susceptibility maps comparison(2), SS-TV-QSM results have obvious image blurring and apparent streaking artifacts, as shown (2, a, i-iv). LN-QSM results suffers image intensity variation across the brain volume and shading artifacts near the brain boundary, as shown (2, b, i-iv). The QSM images of PDF+MEDI show severe image blurring and streaking artifacts, as shown (2, c, i-iv). RESHARP+TKD results suffer from brain erosion, image blurring, and streaking artifacts, as shown (2, d, i-iv). Compared with other methods, QSMAllNet results (2, e, i-iv) show best image quality with high image sharpness and negligible streaking artifacts. 

\begin{figure}[ht]
\begin{center}
\includegraphics[width=14cm]{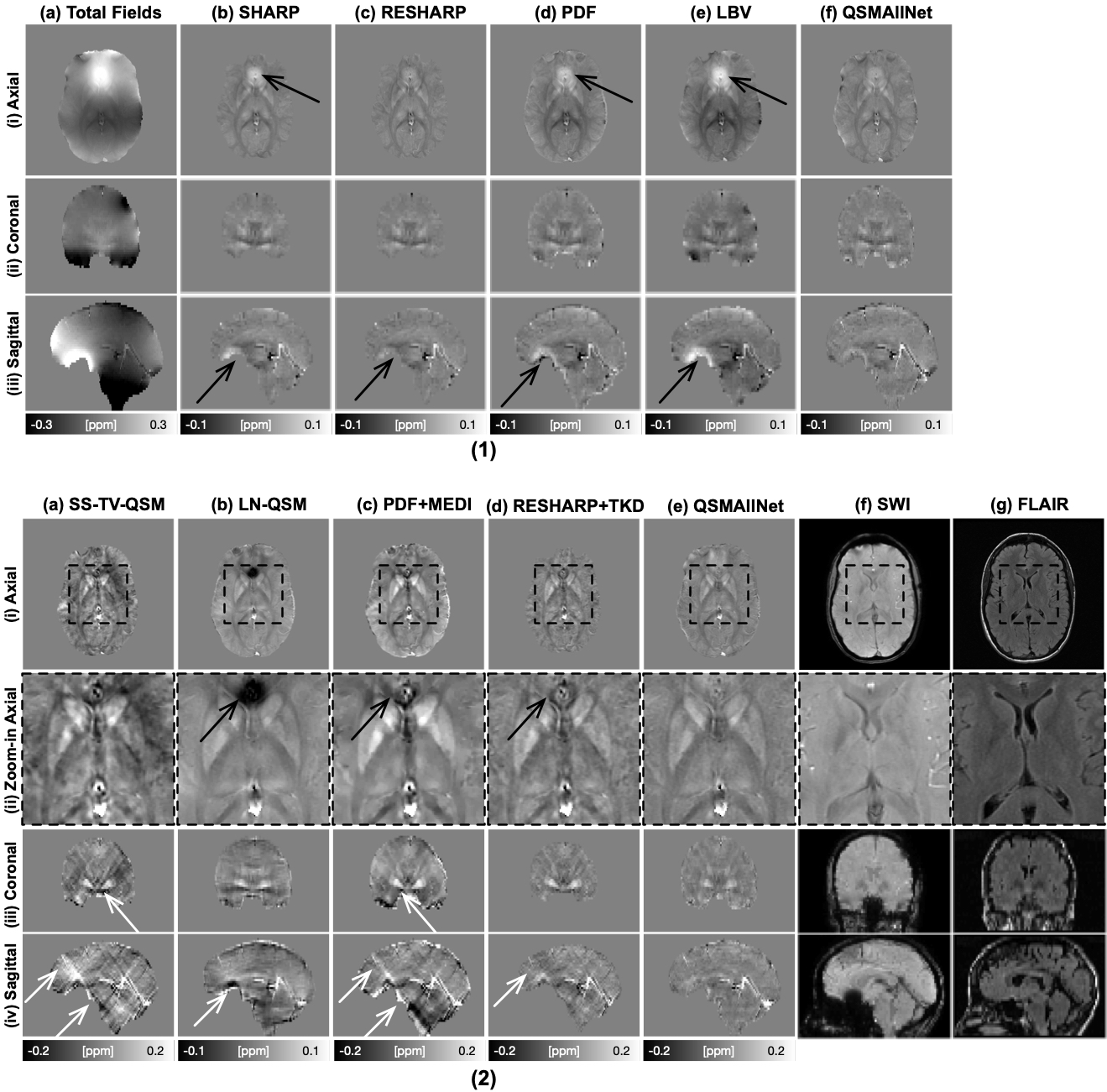}
\caption{Total fields and QSM background removal results (1), susceptibility images, SWI images and FLAIR images (2) on a xx-year-old subject with xxx. In (1), residual background fields are clearly visible in SHARP, RESHARP, PDF and LBV (b-d) results in axial and saggital views. SHARP and RESHARP results (b-c) have brain erosion. LBV and PDF have shading artifacts in the tissue fields. QSMAllNet results show better background field removal. From the susceptibility maps (2), SS-TV-QSM, LN-QSM, PDF+MEDI, and RESHARP+TKD have susceptibility large estimation errors, especially streaking artifacts and shading artifacts ( white arrows). Based on visual comparison, QSMAllNet can produce improved local field and susceptibility estimation.
\label{p582}}
\end{center}
\end{figure}

In Fig.~\ref{p582}, background removal results and susceptibility images of a 37-year-old subject with meningioma and multiple sclerosis is illustrated. Residual background field is visible in SHARP, RESHARP, PDF, and LBV results (1, a-d, i-iv). PDF and LBV results show obvious shading artifacts (1, c-d, i-iv). From the QSM images, SS-TV-QSM, PDF+MEDI, and RESHARP+TKD suffers severe image blurring and streaking artifacts, as shown (2, a, c, d, i-iv). LN-QSM shows strong shading artifacts, as shown (2, b, i-iv). With comparison to other methods, QSMAllNet results (2, e, i-iv) showed improved image sharpness, clear tissue structures, and no image artifacts.

\begin{figure}[ht]
\begin{center}
\includegraphics[width=14cm]{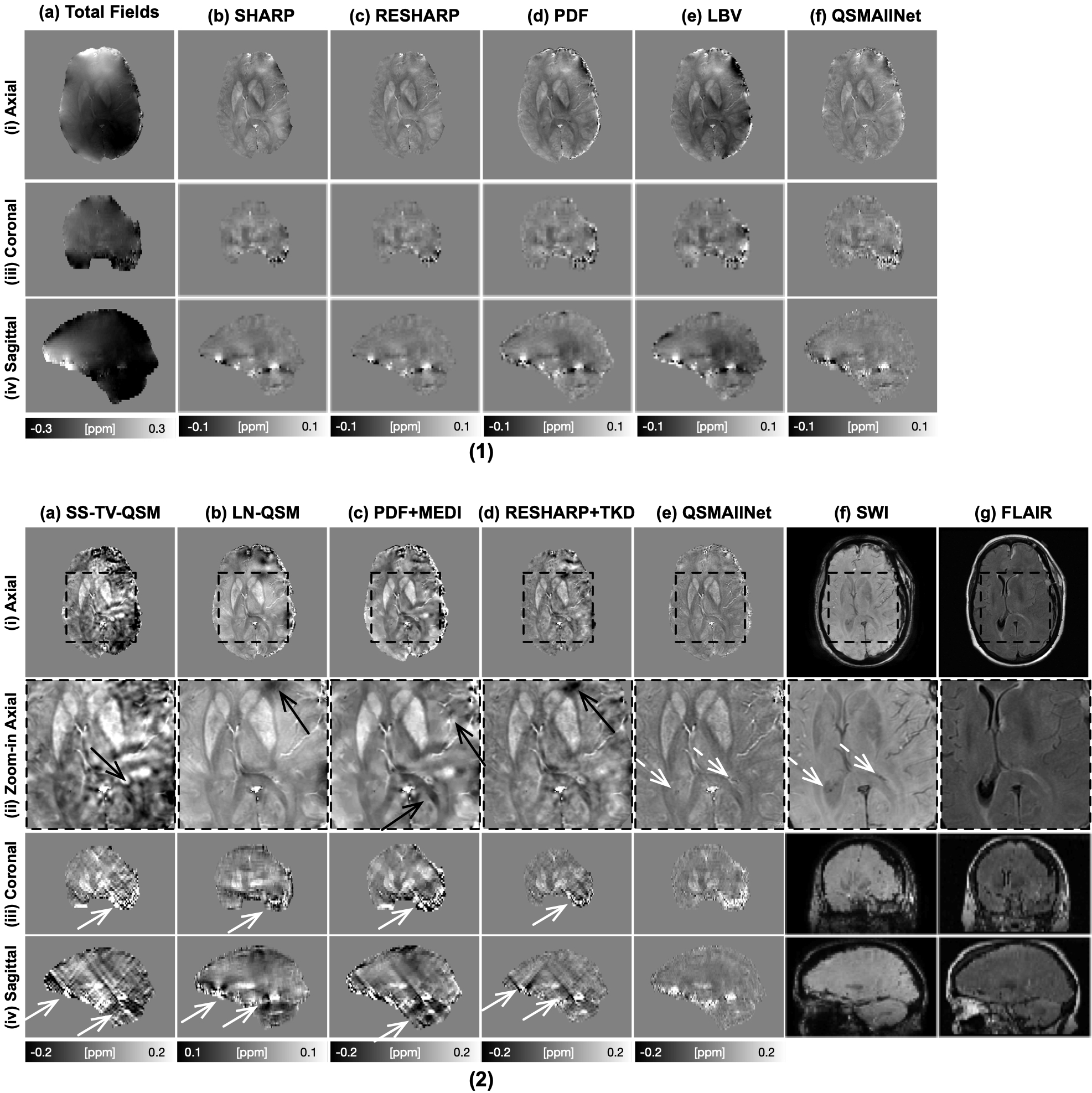}
\caption{Total fields and QSM background removal results (1), susceptibility images, SWI images and FLAIR images (2) on a 56-year-old subject with hemorrhagic intracranial metastases. In (1), residual background fields are clearly visible in SHARP, RESHARP, PDF and LBV (b-d) results in axial and saggital views. SHARP and RESHARP results (b-c) have brain erosion. LBV and PDF have shading artifacts in the tissue fields. QSMAllNet results show better background field removal. From the susceptibility maps (2), SS-TV-QSM, LN-QSM, PDF+MEDI, and RESHARP+TKD have susceptibility large estimation errors, especially streaking artifacts and shading artifacts (white arrows). Two small calcification is dark/hypointense on SWI image and diamagnetic on QSM images (white dash arrows). Based on visual comparison, QSMAllNet can produce improved local field and susceptibility estimation.
\label{p1619}}
\end{center}
\end{figure}

In Fig.~\ref{p1619}, a 56-year-old subject with hemorrhagic intracranial metastases is illustrated. PDF and LBV results show obvious shading artifacts (1, d-e, i-iii). QSMAllNet shows invisible artifacts in the local field. From the QSM images, SS-TV-QSM, PDF+MEDI, and RESHARP+TKD suffers severe shading artifacts and streaking artifacts, as shown (2, a, c, d, i-iv). LN-QSM shows strong shading artifacts, as shown (2, b, i-iv). With comparison to other methods, QSMAllNet results (2, e, i-iv) showed improved image sharpness, clear tissue structures, and no shading and streaking artifacts around bleeding regions. Besides, two small calcifications are dark/hypointense on SWI images and diamagnetic on QSM images (white dash arrows, 2, e-f, ii).

\begin{figure}[ht]
\begin{center}
\includegraphics[width=14cm]{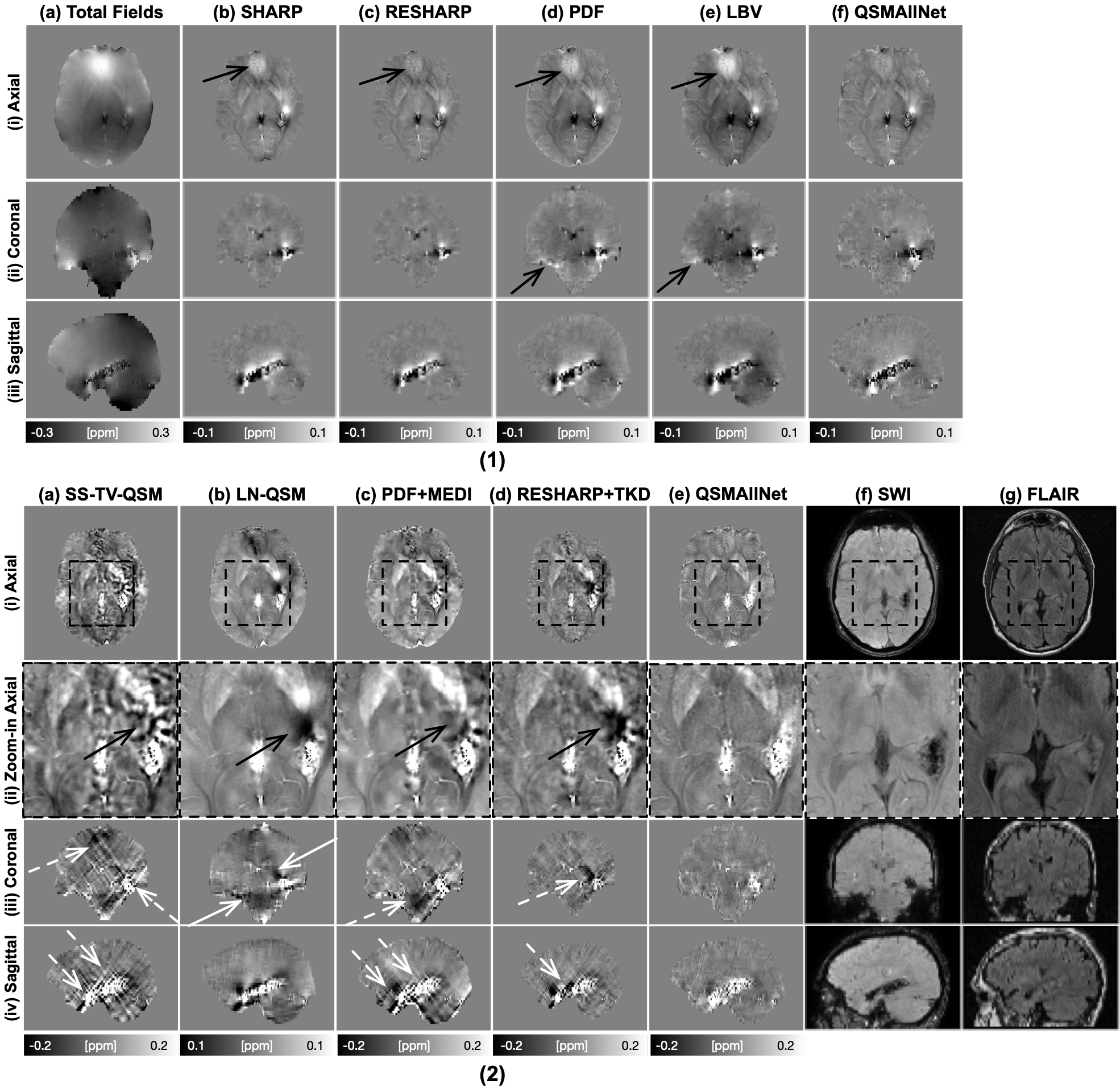}
\caption{Total fields and QSM background removal results (1), susceptibility images, SWI images and FLAIR images (2) on a 28-year-old subject with left mesial temporal lesion and Neurofibromatosis Type-1. In (1), residual background fields are clearly visible in SHARP, RESHARP, PDF and LBV (b-d) results in axial and saggital views. SHARP and RESHARP results (b-c) have brain erosion. LBV and PDF have shading artifacts in the tissue fields. QSMAllNet results show better background field removal. From the susceptibility maps (2), SS-TV-QSM, LN-QSM, PDF+MEDI, and RESHARP+TKD have susceptibility large estimation errors, especially streaking artifacts and shading artifacts (white arrows). Based on visual comparison, QSMAllNet can produce improved local field and susceptibility estimation.
\label{p1388}}
\end{center}
\end{figure}

In Fig.~\ref{p1388}, background removal results and susceptibility images of a 28-year-old subject with left mesial temporal lesion and Neurofibromatosis Type-1 is illustrated. Residual background field is visibile in SHARP, RESHARP, PDF, and LBV results (black arrows, 1, b-e, i-iii). PDF and LBV results show obvious shading artifacts (1, d-e, i-iii). QSMAllNet shows invisible artifacts in the local field. From the QSM images, SS-TV-QSM, PDF+MEDI, and RESHARP+TKD suffers severe shading artifacts and streaking artifacts, as shown (2, a, c, d, i-iv). LN-QSM shows strong shading artifacts, as shown (2, b, i-iv). With comparison to other methods, QSMAllNet results (2, e, i-iv) showed improved image sharpness, clear tissue structures, and no shading and streaking artifacts around bleeding regions. 

\begin{figure}[ht]
\begin{center}
\includegraphics[width=14cm]{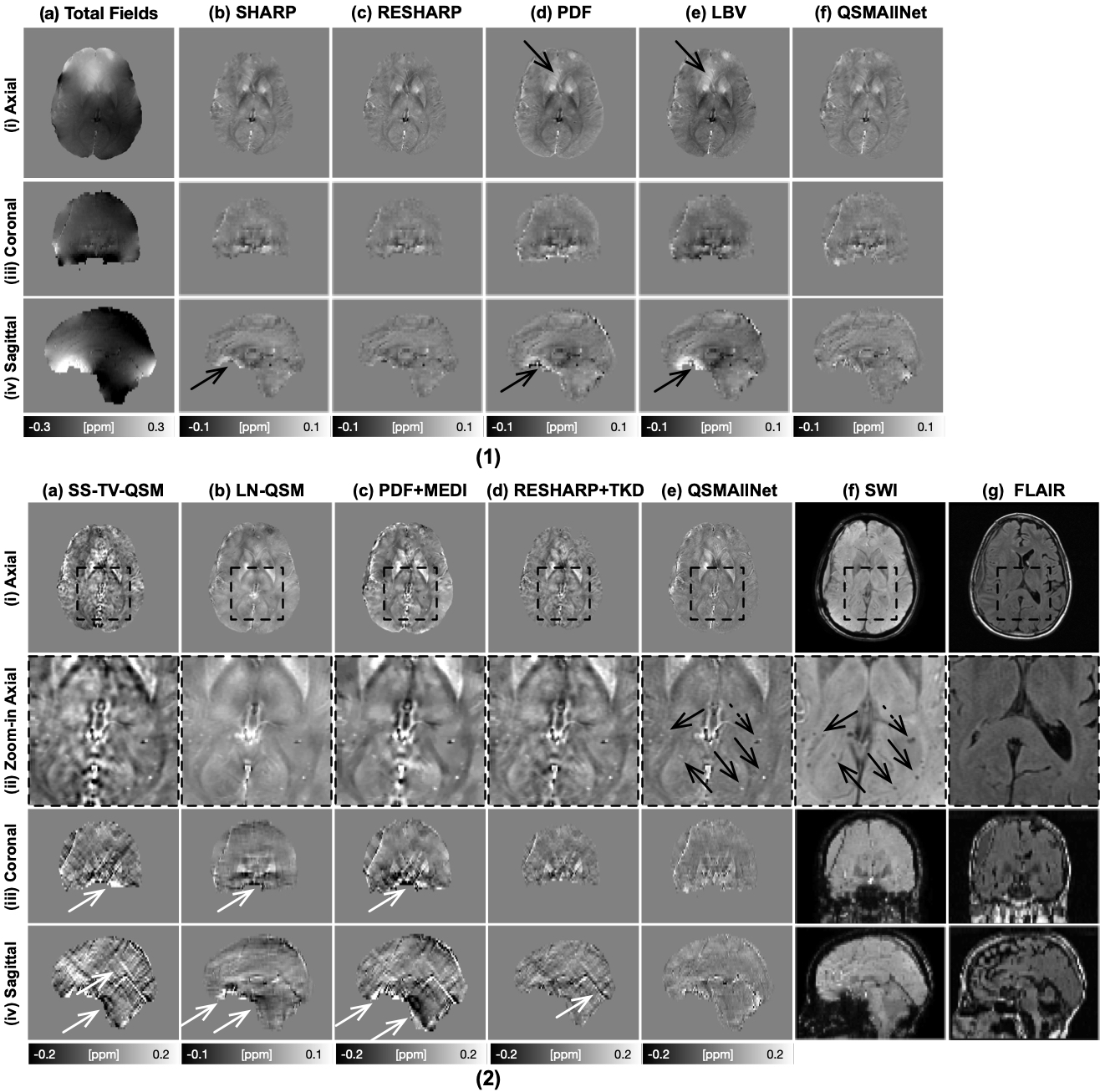}
\caption{Total fields and QSM background removal results (1), susceptibility images, SWI images and FLAIR images (2) on 34-year-old subject with subdural fluid collection and history of meningioma resection. In (1), residual background fields are clearly visible in SHARP, RESHARP, PDF and LBV (b-d) results in axial and saggital views. LBV and PDF have shading artifacts in the tissue fields. From the susceptibility maps (2), SS-TV-QSM, LN-QSM, PDF+MEDI, and RESHARP+TKD have susceptibility large estimation errors, especially streaking artifacts and shading artifacts (white arrows). One small calcification is dark/hypointense on SWI image and diamagnetic on QSM images (black dash arrow). The micorbleeds are dark/hypointense on SWI and QSM images (black solid arrows). Based on visual comparison, QSMAllNet can produce improved local field and susceptibility estimation.
\label{p597}}
\end{center}
\end{figure}

In Fig.~\ref{p597}, a 34-year-old subject with subdural fluid collection and history of meningioma resection is illustrated. Residual background field is visibile in SHARP, PDF, and LBV results (black arrows, 1, b, d, e, i-iii). PDF and LBV results show obvious shading artifacts (1, d-e, i-iii). From the QSM images, SS-TV-QSM, PDF+MEDI, and RESHARP+TKD suffers severe shading artifacts and streaking artifacts, as shown (2, a, c, d, i-iv). LN-QSM shows strong shading artifacts, as shown (2, b, i-iv). One small calcification is dark/hypointense on SWI image and diamagnetic on QSM images (black dash arrow, 2, e-f, ii). The micorbleeds are dark/hypointense on SWI and QSM images (black solid arrows, 2, e-f, ii). Based on visual comparison, QSMAllNet can produce improved local field and susceptibility estimation.

\section{Discussion} 

In this work, QSMAllNet was evaluated on synthetic data, QSM challenge dataset, and clinical dataset. Quantitative metrics and visual assessment demonstrated that QSMAllNet outperformed existing background field removal methods and QSM reconstruction approaches. With comparison with SHARP and RESHARP, the proposed method can preserve the brain tissue without brain erosion. Compared with PDF, the proposed method have less errors on brain boundary and no shading artifacts in local field. With comparision with LBV, QSMAllNet results have no visible residual background field and shading artifacts. 

Compared with single-step and two-step QSM reconstruction approach, QSMAllNet results have better image quality and less quantification errors. Compared with two-step QSM methods, PDF+MEDI and RESHARP+TKD, QSMAllNet can not only speed up the QSM processing but also eliminate the background field removal error propagation into susceptibility estimation. Besides, the proposed methods overcome the constraints of existing background removal approaches, such as brain erosion, parameter tuning and image artifacts. From the clinical SWI examples, neither existing single-step QSM or two-step QSM approaches can produce clinical diagnostic QSM images, suffering from over-smoothing, streaking artifacts, shading artifacts, and large susceptibility quantification errors close to intracerebral hemorrhage regions. The proposed QSMAllNet can overcome the limitations and greatly improve susceptibility quantification accuracy. 

The presented QSMAllNet introduces several important innovations. First, it is trained using synthetic data generated using in-vivo QSM data and physical models, which overcome the bottleneck to obtain labels. Second, it performs whole brain high-resolution QSM background field removal and inversion using a neural network. Compared with using two neural networks or two sub-networks in one large networks to infer local field and susceptibility maps sequentially, our network has simple architecture and less memory burden for training. Third, it utilizes dilated convolution layers to increase the receptive field for non-local susceptibility estimation. From our neural network, the local field and susceptibility map share the same latent space, which needs further work to interpret the neural network.

This feasibility study has also demonstrated the ability to use existing clinical SWI raw data to reconstruct QSM for clinical utility. This offers the possibility of QSM use in clinical operation without any additional scans beyond current standard protocols. Combining SWI magnitude and QSM estimation images may offer new diagnostic capabilities to assist radiological interpretation.

\section{Conclusion} 
In summary, a deep-learning-based approach to perform background field removal and QSM total field inversion have been demonstrated. It can substantially improve accuracy of brain background field removal and susceptibility estimation. The clinical data demonstrated the advantages of QSMAllNet to perform background field removal and QSM reconstruction. This capability opens up a wide array of QSM investigations using clinically acquired SWI data to derive QSM maps across a host of neuroimaging indications.

\bibliographystyle{unsrt}
\bibliography{reference.bib}

\newpage     
	     

\begin{table}[!ht]
\centering
\caption{Numerical measures of QSM background field removal quality on 100 synthetic data.
\label{table_sim_data_rdf}
\vspace*{2ex}}
\begin{tabular}{|c|c|c|c|c|c|}
\hline
\multicolumn{1}{|c}{} & 
\multicolumn{1}{|c}{SHARP} & 
\multicolumn{1}{|c}{RESHARP} &
\multicolumn{1}{|c}{PDF} &
\multicolumn{1}{|c}{LBV} & 
\multicolumn{1}{|c|}{QSMAllNet}\\
\hline 
{{RMSE} ($\%$)} & 
{$52.0\pm4.1$} & 
{$53.5\pm3.7$} & 
{$48.3\pm6.0$} & 
{$29.9\pm3.8$} & 
{\bf{$15.9\pm3.4$}} \\
\hline
{{HFEN} ($\%$)} & 
{$46.9\pm4.2$} & 
{$48.9\pm4.2$} & 
{$37.0\pm4.4$} & 
{$24.9\pm3.7$} & 
{\bf{$14.5\pm3.5$}} \\
\hline 
\multicolumn{1}{|c}{{SSIM} (0-1)} & 
\multicolumn{1}{|c}{$0.734\pm0.016$} & 
\multicolumn{1}{|c}{$0.732\pm0.016$} & 
\multicolumn{1}{|c}{$0.994\pm0.002$} & 
\multicolumn{1}{|c}{$0.858\pm0.008$} & 
\multicolumn{1}{|c|}{\bf{$0.998\pm0.001$}} \\
\hline 
\end{tabular}
\end{table}

\begin{table}[!ht]
\centering
\caption{Numerical measures of QSM reconstruction quality on 100 synthetic data.
\label{table_sim_data_qsm}}
\begin{tabular}{|c|c|c|c|c|c|}
\hline
\multicolumn{1}{|c}{} & 
\multicolumn{1}{|c}{SS-TV-QSM} & 
\multicolumn{1}{|c}{LN-QSM} &
\multicolumn{1}{|c}{RESHARP+TKD} & 
\multicolumn{1}{|c|}{QSMAllNet}\\
\hline 
{{RMSE} ($\%$)} & 
{$43.9\pm4.5$} & 
{$76.6\pm7.6$} & 
{$61.8\pm4.0$} & 
{\bf{$36.2\pm3.2$}} \\
\hline
{{HFEN} ($\%$)} & 
{$38.1\pm3.8$} & 
{$75.2\pm7.9$} & 
{$58.9\pm4.6$} & 
{\bf{$34.2\pm3.5$}} \\
\hline 
\multicolumn{1}{|c}{{SSIM} (0-1)} & 
\multicolumn{1}{|c}{$0.784\pm0.010$} & 
\multicolumn{1}{|c}{$0.897\pm0.027$} & 
\multicolumn{1}{|c}{$0.723\pm0.017$} & 
\multicolumn{1}{|c|}{\bf{$0.976\pm0.006$}} \\
\hline 
\end{tabular}
\end{table}

\begin{figure}[ht]
\begin{center}
\includegraphics[width=14cm]{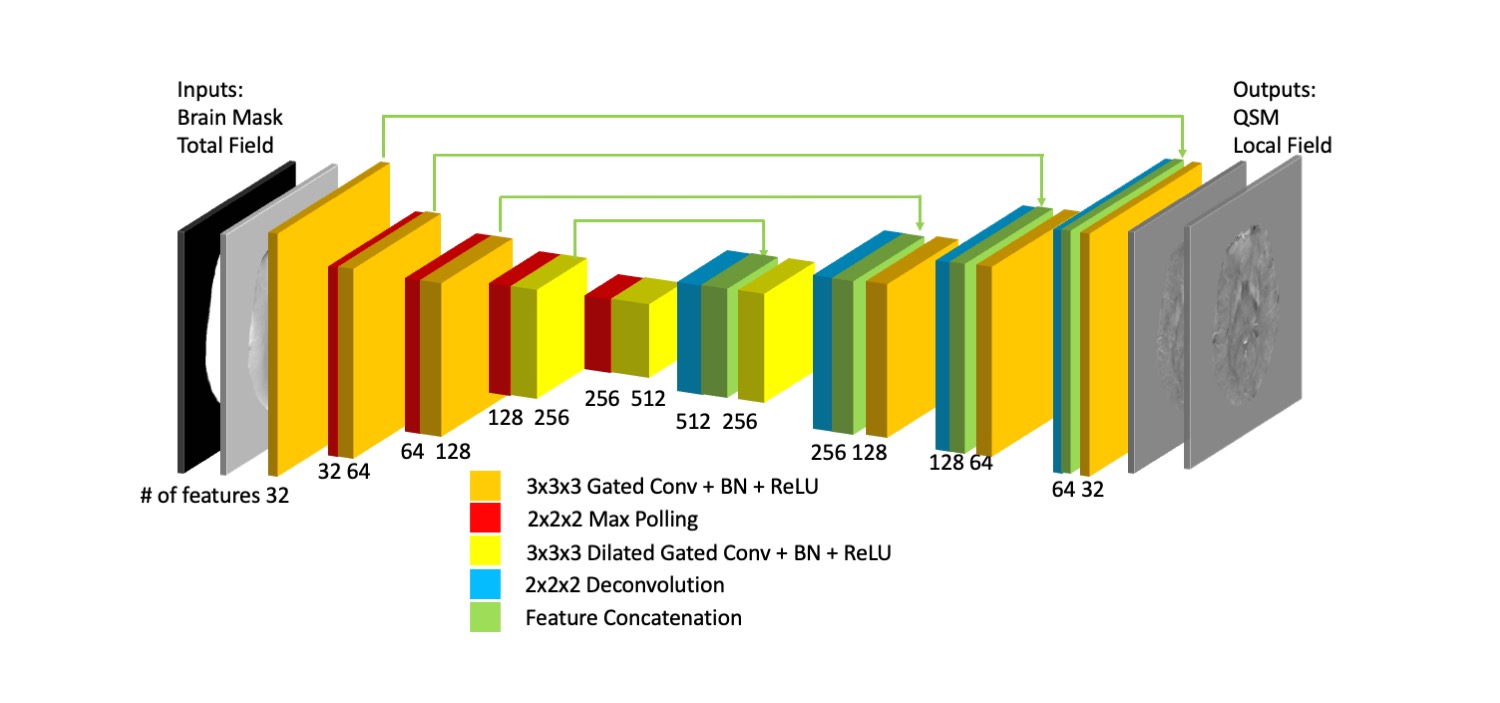}
\caption{Network structure of QSMAllNet. A 3D encoder-decoder architecture was designed with 6 gated convolutional layers (kernel size 3x3x3, dilated rate 1x1x1), 3 gated convolutional layers (kernel size 3x3x3, dilated rate 2x2x2), 4 max pooling layer (pool size 2x2x2, stride size 2x2x2), 1 non-local block, 4 deconvolution layers (kernel size 2x2x2, stride size 2x2x2), 9 normalization layers, 5 feature concatenations, and 1 convolutional layer (kernel size 3x3x3, linear activation). 
\label{fig_ssQSMNet}}
\end{center}
\end{figure}

\end{document}